\documentclass{kapproc} 
\usepackage{graphicx,longtable}

\kluwerbib

\newcommand \lta {\mathrel{\vcenter
     {\hbox{$<$}\nointerlineskip\hbox{$\sim$}}}}
\newcommand \gta {\mathrel{\vcenter
     {\hbox{$>$}\nointerlineskip\hbox{$\sim$}}}}
\newcommand \m {M$_\odot$}

\newcommand\grb{$\gamma$-ray burst}
\newcommand\grbs{$\gamma$-ray bursts}
\newcommand\apj{ApJ}
\newcommand\apjl{ApJL}
\newcommand\apjs{ApJ Supp.}
\newcommand\apss{Ap \& Sp. Sci.}
\newcommand\aap{A \& A}
\newcommand\nat{Nature}
\newcommand\mnras{MNRAS}

\begin{document}

\articletitle{Magnetic Fields in Supernovae}


\author{Shizuka Akiyama}
\affil{Astronomy department\\
University of Texas at Austin\\
1 University Station C1400\\
Austin, TX 78712}
\email{shizuka@astro.as.utexas.edu}

\author{J. Craig Wheeler}
\affil{Astronomy department\\
University of Texas at Austin\\
1 University Station C1400\\
Austin, TX 78712}
\email{wheel@astro.as.utexas.edu}

\begin{abstract}

A relatively modest value of the initial rotation of the iron core, 
a period of $\sim$ 6 -- 31 s, will give a very rapidly rotating 
protoneutron star  and hence
strong differential rotation with respect to the infalling matter. Under
these conditions, a seed field is expected to be amplified by the MRI and
to grow exponentially.  
Exponential growth of the field on the time scale $\Omega^{-1}$ by the
magnetorotational instability (MRI) will dominate the linear growth
process of field line ``wrapping" with the same characteristic time. 
The shear is strongest at the boundary of the
newly formed protoneutron star.  Modest initial rotation velocities of the
iron core result in sub--Keplerian rotation and a sub--equipartition
magnetic field that nevertheless produce substantial MHD luminosity and
hoop stresses: saturation fields of order $10^{15}$ -- $10^{16}$ G develop
$\sim$ 300 msec after bounce with an associated MHD luminosity of $\sim
10^{49}$ -- $10^{53}$ erg s$^{-1}$.  Bi-polar flows driven by this MHD
power can affect or even cause the explosions associated with
core-collapse supernovae. If the initial rotation is too slow, then there
will not be enough rotational energy to power the supernova despite the
high luminosities.  The MRI should be active and may qualitatively
alter the flow if a black hole forms directly or after a fall-back delay.

\end{abstract}


\section*{Introduction}

Accumulating evidence shows that core collapse supernovae are distinctly
and significantly asymmetric.  A number of supernova remnants show
intrinsic ``bilateral" structure (Dubner et al. 2002).  Jet and counter  
jet structures have been mapped for Cas A in the optical (Fesen \&   
Gunderson 1996; Fesen 2001; and references therein), and the intermediate 
mass elements are ejected in a roughly toroidal configuration (Hughes et
al. 2000;  Hwang et al. 2000; Willingale et al. 2002).  The debris of    
SN~1987A has an axis that roughly aligns with the small axis of the rings 
(Pun et al. 2001; Wang et al. 2002).  Spectropolarimetry shows that
substantial asymmetry is ubiquitous in core-collapse supernovae, and that 
a significant fraction of core-collapse supernovae have a bi-polar
structure (Wang et al. 1996, 2001).  The strength of the asymmetry
observed with polarimetry is higher (several \%) in supernovae of Type Ib
and Ic that represent exploding bare non-degenerate cores (Wang et al.
2001).  The degree of asymmetry also rises as a function of time for Type
II supernovae (from $\lta$ 1\% to $\gta$ 1\%) as the ejecta expand and the
photosphere recedes (Wang et al.  2001; Leonard et al. 2000, 2001).  Both
of these trends suggest that it is the core collapse mechanism itself that
is responsible for the asymmetry.

Two possibilities are being actively explored to account for the observed
asymmetries.  One is associated with the rotational effect on convection
(Fryer \& Heger 2000), and another is due to the effect of jets (Khokhlov 
et al. 1999; Wheeler et al. 2000; Wheeler, Meier \& Wilson 2002).  Jet 
calculations have established that non-relativistic axial jets of energy  
of order $10^{51}$ erg originating within the collapsed core can initiate
a bi-polar asymmetric supernova explosion that is consistent with the
spectropolarimetry (Khokhlov et al. 1999; Khokhlov \& H\"oflich 2001;  
H\"oflich et al. 2001).  The result is that heavy elements (e.g. O, Ca)
are characteristically ejected in tori along the equator.  Iron, silicon
and other heavy elements in Cas A are distributed in this way (Hwang et
al. 2000), and there is some evidence for this distribution in SN 1987A
(Wang et al. 2002).  Radioactive matter ejected in the jets can alter the
ionization structure and hence the shape of the photosphere of the
envelope even if the density structure is spherically symmetric (H\"oflich
et al. 2001).  This will generate a finite polarization, even though the
density distribution is spherical and the jets are stopped deep within the
star and may account for the early polarization observed in Type II
supernovae (Leonard et al. 2000; Wang et al. 2001).  If one of the pair of
axial jets is somewhat stronger than the other, jets can, in principle,
also account for pulsar runaway velocities that are parallel to the spin 
axis (Helfand et al. 2001, and references therein).  While a combination
of neutrino--induced and jet--induced explosion may prove necessary for  
complete understanding of core-collapse explosions, jets of the strength
computed by Khokhlov et al. (1999) are sufficient for supernova   
explosions.

Immediately after the discovery of pulsars there were suggestions that   
rotation and magnetic fields could be a significant factor in the
explosion mechanism (Ostriker \& Gunn 1971; Bisnovatyi-Kogan 1971;
Bisnovatyi-Kogan \& Ruzmaikin 1976; Kundt 1976).  Typical dipole fields of
$10^{12}$ G and rotation periods of several to several tens of
milliseconds yield electrodynamic power of $\sim 10^{44-45}$ erg s$^{-1}$
that is insufficient to produce a strong explosion.  The evidence for
asymmetries and the possibility that bi-polar flows or jets can account 
for the observations suggest that this issue must be revisited.  The fact
that pulsars like those in the Crab and Vela remnants have jet-like   
protrusions (Weisskopf et al. 2000; Helfand et al. 2001) also encourages
this line of thought.  The present-day jets in young pulsars may be
vestiges of much more powerful MHD jets that occurred when the pulsar was
born.  The transient values of the magnetic field and rotation could have
greatly exceeded those observed today.  Tapping that energy to power the
explosion could be the very mechanism that results in the modest values of
rotation and field the pulsars display after the ejecta have dispersed.


Possible physical mechanisms for inducing axial flows, asymmetric
supernovae, and related phenomena driven by magnetorotational effects were
considered by Wheeler et al. (2000), who focused on the effect of the
resulting net dipole field, and Wheeler, Meier, \& Wilson (2002) explored
the capacity of the toroidal field to generate axial jets by analogy with
magneto-centrifugal models of jets in AGN (Koide et al. 2000; and
references therein).  Wheeler et al. (2000), Wheeler, Meier, \& Wilson
(2002), and, indeed, all previous work considered only amplification of
the field by ``wrapping," a process that increases the field linearly, and
hence rather slowly in time.  In addition, reconnection might limit the
field before it can be wrapped the thousands of times necessary to be    
interesting.  Akiyama et al. (2002) considerd the effects of magnetic
shearing, the magnetorotational instability (MRI; Balbus \& Hawley 1991,  
1998), on the strongly shearing environment that must exist in a nascent
neutron star.  This instability is expected to lead to the rapid
exponential growth of the magnetic field with characteristic time scale of
order the rotational period.  While this instability has been widely   
explored in the context of accretion disks, this was the first time it has
been applied to core collapse.  This instability must inevitably occur in
core collapse and is likely to be the dominant mechanism for the
production of magnetic flux in the context of core collapse.  This process
has the capacity to produce fields that are sufficiently strong to affect,
if not cause, the explosion.

\section{The Magneto-Rotational Instability}

Akiyama et al. (2002) simulated the collapse of a model iron core of a 15
\m\ progenitor with a one-dimensional flux-limited diffusion code
(Myra et al. 1987).  The evolution of the angular velocity profile
$\Omega(r)$ was computed using the radial density profiles produced by the
core collapse code assuming that the specific angular momentum of a given
shell is constant. The magnetic field was obtained using the resulting
$\rho(r)$, $\Omega(r)$, and $d\Omega/dr$ profiles according to the theory
of the MRI.  The MHD luminosity and hoop stress were estimated from the
resulting magnetic field.

\subsection{Angular Velocity Profile}

The 15 \m\ model of \cite{HLW00} attains an angular velocity of 10 rad
s$^{-1}$ (see their Fig. 8) in the center of the iron core at the
precollapse stage.  Their simulations did not include the effects of a
magnetic field.  It is possible that the iron core rotates slower if the
effect of magnetic braking is included (Spruit \& Phinney 1998;  Heger \&
Woosley 2002). \cite{FH00} studied the rotational effects on pure
hydrodynamic core collapse explosions with initial velocity profiles
obtained by \cite{HLW00} with a central rotational velocity of 4 rad
s$^{-1}$. Akiyama et al. (2002) adopted the initial angular velocity
profile, $\Omega_{0}(r)$, of \cite{FH00} (hereafter called the FH profile)
as one case to study in addition to an analytic (MM profile)  form
(M\"{o}chmeyer \& M\"{u}ller 1989; Yamada \& Sato 1994; Fryer \& Heger
2000) and solid body profiles.  The adopted profiles, characterized by the
initial central value of the rotational frequency, $\Omega_{0,c}$ had
small enough angular momentum that little departure from spherical
geometry will occur.

It is inevitable that the collapsing core spins up and generates strong
differential rotation for very general choices of the $\Omega_{0}(r)$
profile, since the inner regions collapse larger relative distances than  
the outer regions.  A strong shear must form at the boundary of the 
protoneutron star (PNS).
At bounce, the original homologous core has a positive gradient in
$\Omega(r)$, and about 50 ms after bounce, the density profile is nearly
identical to that of the initial iron core, giving a nearly flat rotation
profile .  After that, the density profile becomes somewhat more centrally
condensed than the original iron core and the rotation profile decreases
monotonically outward even deep within the PNS (Fig. 1, 2).

\begin{figure}[th]
\centering
\includegraphics[width=11.2cm,clip=,angle=0]{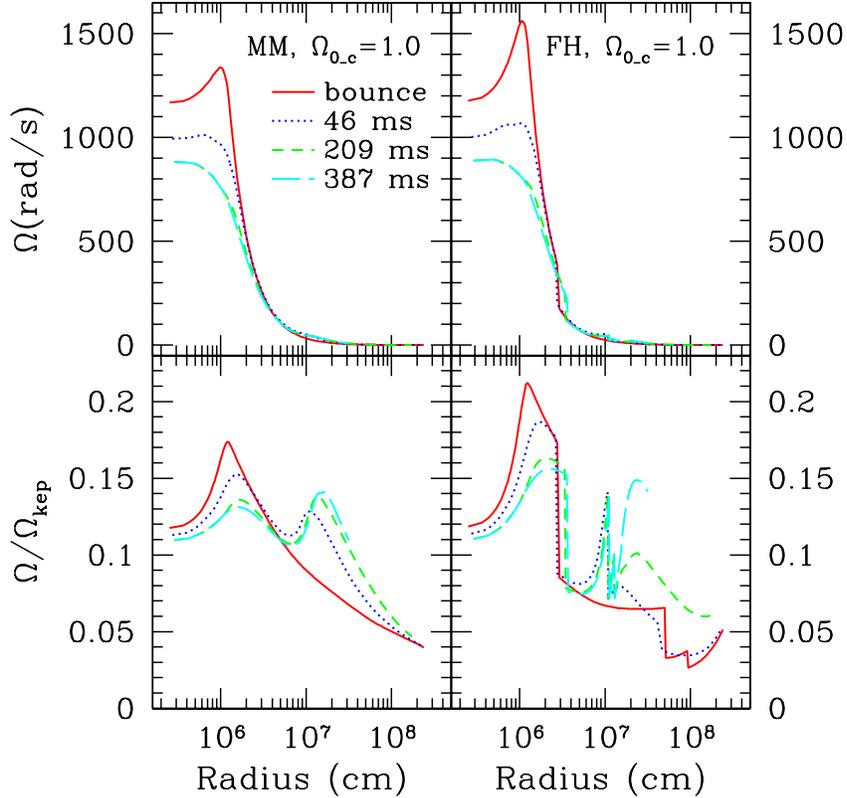}
\caption{Rotational profiles and $\Omega/\Omega_{\rm{kep}}$ for the
initial differential rotation cases (MM and FH) with $\Omega_{0,c}=1.0$
rad s$^{-1}$.  The collapse generates strong differential rotation at the
boundary of the initial homologous core.  The rotation is always 
sub--Keplerian.}
\label{fig1}
\end{figure}

\begin{figure}[th]
\centering
\includegraphics[width=11.2cm,clip=,angle=0]{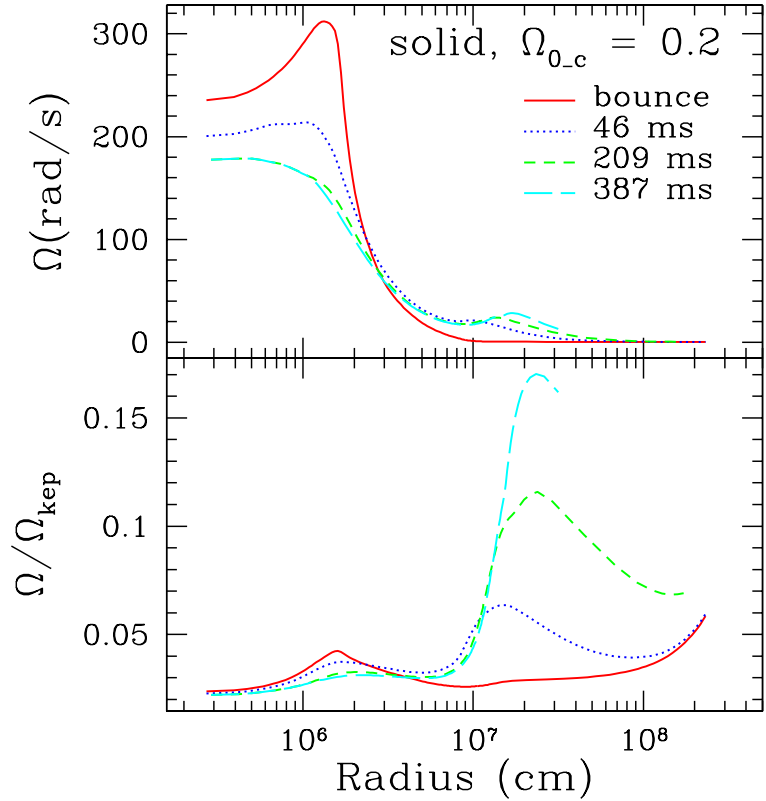}
\caption{Rotational profiles and $\Omega/\Omega_{\rm{kep}}$ for the 
initial solid body rotation case with $\Omega_{0,c}=0.2$ rad s$^{-1}$.  
The rotational profiles are very similar to those of initial differential 
rotation cases.} 
\label{fig2} 
\end{figure}

\cite{rud00} noted that the collapse of a white dwarf to a PNS gives a
positive $\Omega(r)$ gradient since the relativistic degenerate core of
the white dwarf has a steeper density profile than the PNS. The PNS will
thus be relatively more compact for a given central density. There are two
important differences in the calculations of Akiyama et al. (2002).  The  
most critical is that the core collapsing is not in isolation as for the
accretion-induced collapse scenario.  Rather, the PNS forms within the
massive star collapse ambience, and the PNS must be strongly
differentially rotating with respect to the still-infalling matter.  This
will generate a strong shear at the boundary of the PNS that would not
pertain to a collapsing isolated white dwarf.  Another, more subtle,
difference is the equation of state.  The equation of state of a partially
degenerate iron core is not as different as for the highly relativistic
white dwarf collapsing to a non-relativistic neutron star.

\subsection{Magnetic Field}

The MRI generates turbulence in a magnetized rotating fluid body that
amplifies the magnetic field and transfers angular momentum.
The MRI should pertain in this environment and amplify the
magnetic field exponentially and perhaps, in turn, power MHD bi-polar
flow or jets.  Key questions are the amplitude of the magnetic field   
and the effect on the dynamics.

Ignoring entropy gradients, the condition for the
instability of the slow magnetosonic waves in a magnetized, differentially
rotating plasma is (Balbus \& Hawley 1991, 1998):
\begin{equation}
\frac{d\Omega^2}{d\ln{r}} + ({\bf{k \cdot v_{\rm{A}}}})^2 < 0,
\end{equation}
where
\begin{equation}
\label{v_alfven}
v_{\rm{A}} = \frac{B}{\sqrt{4 \pi \rho}}
\end{equation}
is the Alfv\'{e}n velocity.  When the magnetic field is very small, and/or
the wavelength is very long,
$({\bf{k \cdot v_{\rm{A}}}})^2$ is negligible, and the instability
criterion for the MRI is simply that the angular velocity gradient be
negative (Balbus \& Hawley 1991, 1998), i.e.:
\begin{equation}
\frac{d\Omega^2}{d\ln{r}} < 0.
\end{equation}

The growth of the magnetic field associated with the MRI is exponential
with characteristic time scale of order $\Omega^{-1}$. The time scale for
the maximum growing mode is given by (Balbus \& Hawley 1998):
\begin{equation}
\label{growth-time}
\tau_{\rm{max}} = 4 \pi \left| \frac{d\Omega}{d\ln{r}} \right|^{-1}. 
\end{equation}
We thus expect the MRI to dominate any process such as ``wrapping of field
lines'' (cf. Wheeler et al. 2000 and references therein) that only grows  
linearly in time, even if on about the same time scale.  The MRI will also
operate under conditions of moderate rotation that are not sufficient to
compete with the PNS convective time scales to drive the sort of
$\alpha$~--~$\Omega$ dynamo invoked by, e.g., Duncan \& Thompson (1992).
The resulting unstable flow is expected to become non-linear, develop
turbulence, and drive a dynamo that amplifies and sustains the field.

An order of magnitude estimate for the saturation field can be obtained by
equating the shearing length scale $\ell_{\rm{shear}} \sim
dr/d\ln{\Omega}$ to the characteristic mode scale $\ell_{\rm{mode}} \sim
v_{\rm{A}} \cdot (d\Omega/d\ln{r})^{-1}$.  The resulting saturation
magnetic field is
given by:
\begin{equation}
\label{bsat} 
B_{\rm{sat}}^2 \sim 4 \pi \rho r^2 \Omega^2.
\end{equation}
This is the same result as obtained by setting the Alfv\'{e}n velocity
equal to the local rotational velocity, $v_{\rm{A}} = r \Omega$.

The empirical value of the saturation field obtained by the numerical
simulation of \cite{HGB96} is:
\begin{eqnarray}
B_{\rm{sim}} &=& \sqrt{\frac{\rho}{\pi}} r \Omega \nonumber \\
\label{bsim}
&=& \frac{1}{2\pi} \cdot B_{\rm{sat}}.
\end{eqnarray}
This saturation field is achieved after turbulence is fully established,
which takes about 20 rotations following the initial exponential growth
(Hawley et al. 1996).  For conditions of rotation at much less than
Keplerian, these saturation
fields are much
less than the equipartition field for which $B^2/8\pi$ is comparable to
the ambient pressure, i.e. for the current calculations,
$c_{\rm{s}} \gg r\Omega \sim v_{\rm{A}}$.

When a vertical seed
field exists, the maximum unstable growing mode (Balbus \& Hawley 1998) 
implies a
saturation field of:
\begin{equation}
\label{bmax}
B_{\rm{max}}^2 = - 4 \pi \rho \lambda_{\rm{max}}^2 \Omega^2 \cdot
\left[ \frac{1}{8 \pi^2} \left( 1 + \frac{1}{8}
\frac{d\ln{\Omega^2}}{d\ln{r}} \right)
\frac{d\ln{\Omega^2}}{d\ln{r}}\right],
\end{equation}
where $\lambda_{\rm{max}}$ is the wavelength of the maximum growing mode
 which is not
allowed to exceed the local radius $r$.  With $\lambda_{\rm{max}} = r$,
eq. (\ref{bmax})  becomes:
\begin{equation}
\label{bmax2}
B_{\rm{max}}^2 = - B_{\rm{sat}}^2 \cdot \left[ \frac{1}{8 \pi^2} \left( 1
+ \frac{1}{8} \frac{d\ln{\Omega^2}}{d\ln{r}} \right)
\frac{d\ln{\Omega^2}}{d\ln{r}}\right].  
\end{equation}
This expression for the saturation field depends on the shear explicitly
as well as indirectly through the stability criterion.
Note that for the
maximum growing mode the expression for $B_{\rm{max}}^2$ acquires a
negative value when
\begin{eqnarray}
\frac{d\Omega^2}{d\ln{r}} & < & - 8\Omega^2 \mbox{ or }, \nonumber \\
\label{omegalim}
\kappa^2 & < & - 4\Omega^2 < 0,
\end{eqnarray}
where $\kappa$ is the epicyclic frequency:
\begin{equation}
\kappa^2 = \frac{1}{r^3}\frac{d(r^4\Omega^2)}{dr} = 4\Omega^2 +
\frac{d\Omega^2}{d\ln{r}}.
\end{equation}
When eq. (\ref{omegalim}) is true, the epicyclic motion dominates over
the MRI and prevents growth of the perturbation.  Akiyama et al. (2002)
turned off field amplification when this condition arose.  In practice,
the gradient of $\Omega$ may be reduced by mixing due to the epicyclic
motion, and the MRI may eventually be active in a region in which it was
at first suppressed by a strong negative gradient of $\Omega(r)$. Akiyama
et al. (2002) also discuss the situation when the protoneutron star is
convectively unstable.

\subsection{Results}

The shear is the strongest at the boundary of the initial homologous core.  
At bounce the shear is positive inside of the initial homologous core, and
the region is stable against the MRI though convective instability can
destabilize the structure.  The solid body profile possesses a similar
shear profile to the FH and MM profiles.

Even a relatively modest value of $\Omega_{0}$ gives a very rapidly
rotating PNS and hence strong differential rotation with respect to the
infalling matter.  At bounce, the peak of $\Omega/\Omega_{\rm{kep}}$ is at
the boundary of the initial homologous core.  At later times, however, the
peak moves to the second hump which is located inside the stalled shock.  
This hump is at the same location as a maximum in entropy which is caused
by shocked material with higher density.

\begin{figure}[th]
\centering
\includegraphics[width=11.2cm,clip=,angle=0]{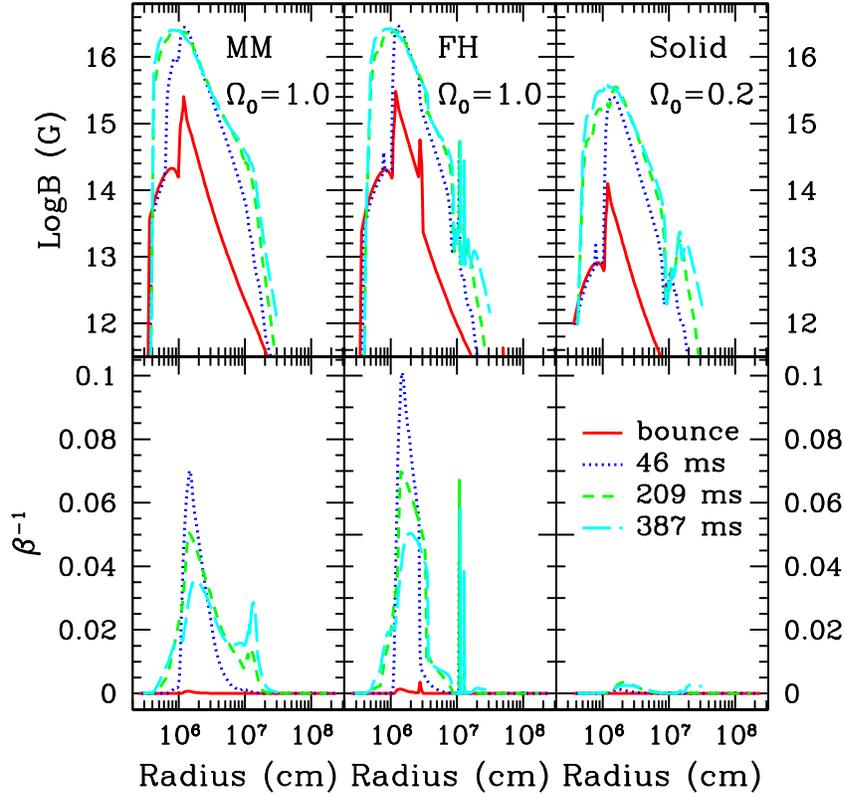}
\caption{Magnetic field that of $B_{\rm{sat}}$ (eq. \ref{bsat}) and the
ratio $\beta^{-1} = P_{\rm{mag}}/P_{\rm{gas}}$ for MM, FH, and solid body
profiles.}
\label{fig3}
\end{figure}

For the given initial rotational profiles, the magnetic fields of eq.  
(\ref{bsat}) and eq. (\ref{bmax2}) are amplified exponentially with 
the time scale of eq. (\ref{growth-time}).  The resulting magnetic field for
$B_{\rm{sat}}$ (eq. \ref{bsat}) and the ratio $\beta^{-1} \equiv
P_{\rm{mag}}/P_{\rm{gas}}$ (where $\beta$ is the conventional $\beta$ in
plasma physics) are presented in Fig. 3.  Given the limitation of the
current calculations, we can only argue that these fields are roughly
representative of what one would expect during core collapse.

For the cases with initial differential rotation, the peak values of the
magnetic field at the end of our calculation at 387 ms after bounce are
$B_{\rm{sat}} = 2.7 \times 10^{16}$ G and $B_{\rm{max}} = 2.5 \times
10^{15}$ G for the FH profile, $B_{\rm{sat}} = 2.5 \times 10^{16}$ G and
$B_{\rm{max}} = 2.5 \times 10^{15}$ G for the MM profile, and
$B_{\rm{sat}} = 3.7 \times 10^{15}$ G and $B_{\rm{max}} = 4.0 \times
10^{14}$ G for the solid body profile. The amplitude of the magnetic field
is remarkably high and above the QED limit ($B_{\rm{Q}} = 4.4 \times
10^{13}$ G), but remains less than equipartition.  For the case of
$B_{\rm{sat}}$, $P_{\rm{mag}}$ is above 10\% of $P_{\rm{gas}}$, and
magnetic buoyancy may limit growth of the magnetic field (Wheeler, 
Meier, \& Wilson 2002).

We expect that the magnetic field generated by the MRI will power MHD
bi-polar outflow. The characteristic power of non-relativistic MHD outflow
is given by
Blandford \& Payne (1982; see also Meier 1999, Wheeler, Meier \&
Wilson 2002):
\begin{equation}
\label{lmhd}
L_{\rm{MHD}} = \frac{B^2r^3\Omega}{2}.
\end{equation}
The outflow carries energy, angular momentum and mass.  Employing this   
characteristic power of a Blandford
\& Payne type MHD outflow, the outflow luminosity
$L_{\rm{MHD}}$ calculated for
the three initial rotational profiles, MM, FH, and solid body.
The profiles of this MHD luminosity mimic those of the magnetic field.
 For the cases with initial
differential profiles, the peaks of the MHD luminosity are at the boundary
of the PNS.

\begin{figure}[th]
\centering
\includegraphics[width=11.2cm,clip=,angle=0]{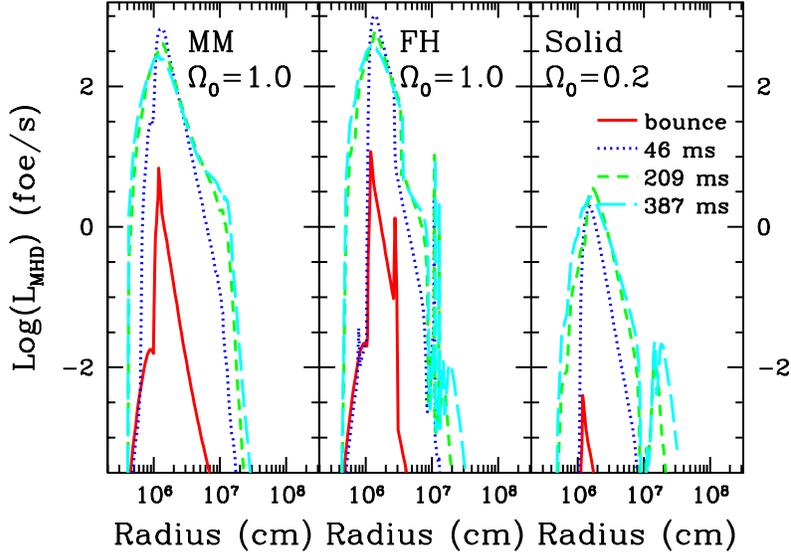}
\caption{MHD jet luminosity in units of 1 foe ($= 10^{51}$ erg)  
corresponding to $B_{\rm{sat}}$.  Sub-Keplerian rotation with MM, FH, and
solid body initial rotation profiles results in high luminosity.}
\label{fig4}
\end{figure}

Our calculations are limited to sub--Keplerian rotation and
sub-equipartition fields, and yet they potentially produce significant
MHD luminosity (Fig. 4): for the saturation field $B_{\rm{sat}}$, the
maximum values 387 ms after bounce are 3.8 $\times 10^{53}$ erg s$^{-1}$
for FH, 2.5 $\times 10^{53}$ erg s$^{-1}$ for MM, and 1.9 $\times 10^{51}$
erg s$^{-1}$ for the initial solid rotation.  For the saturation field
$B_{\rm{max}}$, the peak values of MHD luminosity 387 ms after bounce are
5.5 $\times 10^{51}$ erg s$^{-1}$ for FH, 4.2 $\times 10^{51}$ erg
s$^{-1}$ for MM, and 8.2 $\times 10^{49}$ erg s$^{-1}$ for the initial
solid rotation.  The investigation of how the MHD luminosity can be turned
into a bi-polar flow is left for future work, although we outline some
possibilities in the discussion below.

\section{Discussion and Conclusions}

No one would doubt that the progenitors of core collapse supernovae rotate
and possess some magnetic field.  The question has always been whether
rotation and magnetic fields would be incidental perturbations or a
critical factor in understanding the explosion. Akiyama et al. (2002) have
shown that with plausible rotation from contemporary stellar
evolution calculations and any finite seed field with a component parallel
to the rotation axis, the magnetorotational instability can lead to the
rapid exponential growth of the magnetic field to substantial values on
times of a fraction of a second, comparable to the core collapse time.  
Even a relatively modest value of initial rotation gives a very rapidly
rotating PNS and hence strong differential rotation with respect to the
infalling matter.

This result promises to be robust because the instability condition for
the MRI is basically only that the gradient in angular velocity be
negative.  This condition is broadly satisfied in core collapse
environments.  Rotation can weaken supernova explosions without magnetic
field (Fryer \& Heger 2000); on the other hand, rotational energy can be
converted to magnetic energy that can power MHD bi-polar flow that may
promote supernova explosions.  {\it The implication is that rotation and
magnetic fields cannot be ignored in the core collapse context.}

As expected, the shear and hence the saturation fields are often highest
at the boundary of the PNS where strong MHD activity is anticipated. Even
artificially limiting the post-collapse rotation to sub--Keplerian values
as done by Akiyama et al. (2002), we find fields in excess of $10^{15}$ G
near the boundary of the neutron star are produced.  While this field
strength is sub--equipartition, the implied MHD luminosities are of order
$10^{52}$ erg s$^{-1}$.  This is a substantial luminosity and could,
alone, power a supernova explosion if sustained for a sufficiently long
time, a fraction of a second.  As pointed out by Wheeler, Meier \& Wilson
(2002), the fields do not have to be comparable to equipartition to be
important because they can catalyze the conversion of the large reservoir
of rotational energy into buoyant, bi-polar MHD flow.  Higher rates of
initial rotation that are within the bounds of the evolutionary
calculations could lead to even larger post-collapse rotation and even
larger magnetic fields.  If the initial rotation of the iron core proves
to be substantially lower than we have explored here, then the MRI would
be of little consequence to the explosion.  The MHD luminosities derived
here are comparable to the typical neutrino luminosities derived from core
collapse, $\sim 10^{52}$ erg~s$^{-1}$. One important difference is that
the matter beyond the PNS is increasingly transparent to this neutrino
luminosity, whereas the MHD power is deposited locally in the plasma.
Another difference is that the neutrino luminosity is basically radial so
it resists the inward fall of the collapse, the very source of the
neutrino luminosity itself.  In contrast, hoop stresses associated with
the magnetic field (see below) will tend to pull inward and force matter
selectively up the rotation axis.

We note that for complete self-consistency, one should apply the MRI to
the evolution of rotating stars where even a weak field renders the
H{\o}iland dynamical stability criterion ``all but useless" in the words
of Balbus \& Hawley (1998). Recent calculations by \cite{HW02} based on a
prescription for magnetic viscosity by \cite{spruit02} yield rapidly
rotating iron cores. \cite{HW02} find PNS rotation rates of 4 to 8 ms,
consistent with the values we have explored here.  Clearly, much more must
be done to understand the magnetorotational evolution of supernova
progenitors.

The configuration of the magnetic field in a precollapse iron core is not
well understood.  In this calculation we have assumed there exists a seed
vertical field to calculate the growth of the field due to the MRI;
however, the MRI can amplify other components of the magnetic field.  The
final configuration of the magnetic field after collapse may be less
uncertain since the system has a strongly preferred direction due to
rotation.  Most of the shear is in the radial direction, so the radial
component is greatly amplified by the MRI and turned into toroidal field
due to differential rotation (Balbus \& Hawley 1998).  The dominant
component is most likely to be the toroidal field.

Another uncertainty is the rotational profile. It is not clear what
profile to use in the PNS, since, we note, even the rotational profile of
the Sun is not well understood.  A full understanding of the rotational
state of a PNS remains a large challenge.

We have assumed various prescriptions for the saturation field.  All are
variations on the theme that, within factors of order 2$\pi$, the
saturation field will be given by the condition $v_{\rm{A}} \sim r
\Omega$. In the numerical disk simulations, about 20 rotations are
required to reach saturation.  The region of maximum shear in these
calculations, around 15 km, typically has an angular velocity of 500 rad
s$^{-1}$ or a period of about 0.013 s.  That means that by the end of the
current calculations at 0.387 s, there have been about 30 rotations.
Although the prescriptions for the growth and saturation fields we use
here are heuristic, this aspect of our results is certainly commensurate
with the numerical simulations of the MRI.

The issues of the saturation field and the nature of astrophysical dynamos
are still vigorously explored. Vishniac \& Cho (2001) conclude that the
MRI has the required properties for a dynamo, anisotropic turbulence in a
shearing flow, to generate both disordered and ordered fields of large
strength.  The saturation limits we have adopted here are consistent with
those found in numerical calculations of the MRI saturation, but this
topic clearly deserves more study.

Both the magnetic pressure and the magnetic viscosity are small for the
sub--Keplerian conditions explored here.  For most cases $\beta^{-1}$ is
less than 0.1 for the conditions we have assumed, (the $B_{\rm{sat}}$ case
for FH with $\Omega_{0,\rm{c}} = 1.0$ pushes this limit), so the direct
dynamical effect of the magnetic field is expected to be small.  The
viscous time scale is $\tau_{\rm{vis}} \sim (\alpha
\Omega)^{-1}(\rm{r/h})^2$, where $\alpha$ is the viscosity parameter and h
is the vertical scale height, with h $\sim$ r for our case.  For a
magnetically-dominated viscosity,
\begin{equation}
\alpha \sim \frac{B_{\rm{r}} B_{\phi}}{4 \pi P} =
\left(\frac{B_{\rm{r}}}{B_{\phi}}\right) \frac{B_{\phi}^2}{4 \pi P} \sim 2 
\left(\frac{B_{\rm{r}}}{B_{\phi}}\right) \beta^{-1}.
\end{equation}
With this expression for $\alpha$, the viscous
time becomes:
\begin{equation}
\tau_{\rm{vis}} \sim \frac{1}{2}\left(\frac{B_{\phi}}{B_{\rm{r}}}\right)
\left(\frac{1}{\beta^{-1}\Omega}\right) \gg \Omega^{-1},
\end{equation}
where the final inequality follows from $B_{\phi} > B_{\rm{r}}$ and
$\beta^{-1} < 1$.  This prescription for viscosity is
reasonable in the absence of convection.  In the portions of the structure
that are convective, the viscosity could be enhanced significantly.

We have not discussed the role of neutrinos here, although the processes
of neutrino loss and de-leptonization are included in our calculation of
the cooling PNS.  It is possible that the neutrino flux affects the
magnetic buoyancy (Thompson \& Murray 2002) and that the magnetic fields
affect the neutrino emissivity (\cite{TD96}) and interactions with the
plasma (Laming 1999).  The time scale for shear viscosity due to neutrino
diffusion is much longer than the times of interest here, although
magnetic fields and turbulence can make it shorter (\cite{gou98}).  The
MRI provides magnetic field and turbulence, so this issue deserves further
study.  In addition to affecting the shear, the neutrino viscosity might
also affect the turbulence needed to make the MRI work.

An obvious imperative is to now understand the behavior of the strong
magnetic fields we believe are likely to be attendant to any core collapse
situation.  The fields will generate strong pressure anisotropies that can
lead to dynamic response even when the magnetic pressure is small compared
to the isotropic ambient gas pressure. As argued in Wheeler, Meier \&
Wilson (2002), a dominant toroidal component is a natural condition to
form a collimated magneto-centrifugal wind, and hence polar flow.  A first
example of driving a polar flow with the MRI is given by Hawley \& Balbus
(2002).

The MRI is expected to yield a combination of large scale and small scale
magnetic fields.  A key ingredient to force flow up the axis and to
collimate it is the hoop stress from the resulting field.  Hoop stresses and
other aspects of the strongly anisotropic Maxwell stress tensor will tend
to lead to enhanced flow inward on the equator and up the axis, thus
promoting a jet.  These hoop stresses will occur for a field with a large
scale toroidal component, but also in cases with only a small scale, turbulent
field, i.e. when $<B_{\phi}> = 0$ but $<B_{\phi}^2> \ne 0$ (Ogilvie 2001,
Williams 2002).

Akiyama et al. (2002) found that the acceleration implied by the hoop
stresses of the saturation fields, $a_{\rm{hoop}} = B_{\phi}^2/4\pi \rho
r$, was competitive with, and could even exceed, the net acceleration of
the pressure gradient and gravity.  The large scale toroidal field is thus
likely to affect the dynamics by accelerating matter inward along
cylindrical radii. The flow, thus compressed, is likely to be channeled up
the rotation axes to begin the bi-polar flow that will be further
accelerated by hoop and torsional stresses from the field, the ``spring
and fling" outlined in Wheeler, Meier \& Wilson (2002).

While there has been some excellent work on the generation and propagation
of jets through stars in the context of the ``collapsar" model (Aloy et
al. 1999, 2000; MacFadyen \& Woosley 1999; MacFadyen, Woosley \& Heger
2000; Zhang, Woosley \& MacFadyen 2002)  and for
supernovae (Khokhlov et al. 1999; Khokhlov \& H\"oflich 2001; H\"oflich,
Wang, and Khokhlov 2001), none of this numerical work has taken explicit
account of rotation and magnetic fields in the origin and propagation of
the jet.  The same is true for the associated analytic work on jet
propagation (Tan, Matzner, \& McKee 2001; Matzner 2002; M\'esz\'aros \&
Rees 2001; Ramirez-Ruiz, Celotti \& Rees 2002).

The dynamics of MHD jets may depart substantially from pure hydrodynamical
jets, since they will tend to preserve the flux in the Poynting flow and
be subject to hoop stresses and other magnetic phenomena.  In addition,
reconnection can accelerate the matter (Spruit, Daigne \& Drenkhahn 2001).  
The magnetic forces in the jet may affect the collimation of the jet and
the efficiency with which the surrounding cocoon is heated and expands in
a transverse manner.  A sufficiently strong magnetic field could thus
alter the efficiency with which the propagating jet deposits energy into
the stellar envelope. The magnetic field can also affect the stability of
the jet. Large scale helical fields tend to be unstable to the kink
instability.  Li (2002) has recently argued that the effective hoop
stresses of small scale turbulent fields could collimate magnetic jets and
stabilize the flow against kinking.

Understanding of the role of magnetic fields in supernovae may also shed
light on the production of collimated jets and magnetic field in the more
extreme case of \grbs. One of the outstanding questions associated with
\grbs\ is the origin of the magnetic field that is implicit in all the
modeling of synchrotron emission. The fields deduced from the modeling are
comparable to, but substantially less than, equipartition.  Such fields
cannot arise simply from shock compression of the ambient field of the
ISM.  While some schemes for generating this field in the \grb\ shock have
been proposed (Medvedev \& Loeb 1999), there is no generally-accepted
understanding of the origin of this strong field.

Jets arising from the rotation and magnetic fields of neutron stars are
likely to be important in asymmetric core-collapse supernova explosions
(Wheeler et al. 2000; Wheeler, Meier, \& Wilson 2002; Akiyama et al.  
2002). Rotation and magnetic fields are critical in current models for the
origin of jets in everything from protostars to AGN (Meier et al. 2001)
and are very likely to be involved in the rapidly rotating environment
that must occur if core collapse to a black hole is to produce anything
like a \grb\ in the collapsar scenario.  If the magnetic field plays a
significant role in launching a relativistic \grb\ jet from within a
collapsing star, then the magnetic field may also play a role in the
propagation, collimation, and stability of that jet within and beyond the
star.  These factors have not been considered quantitatively.  If magnetic
flux is carried out of the star in the jet, then the magnetic field
required to explain the observed synchrotron radiation may already be
present and will not have to be generated {\it in situ}.

The MRI can operate under conditions of moderate rotation.
This means that the MRI will be at work even as the disk of material
described by MacFadyen \& Woosley (1999) begins to form and makes a
transition
from a non-Keplerian to quasi-Keplerian flow.   The resulting
unstable flow is expected to become non-linear, develop
turbulence, and drive a dynamo that amplifies and sustains the field.

For a complete understanding of the physics in a core collapse supernova 
explosion, a combination of neutrino--induced and jet--induced explosion
may be required.  Understanding the myriad implications of this statement
will be a rich exploration.

\acknowledgments This work was supported in part by
NASA Grant NAG59302 and NSF Grant AST-0098644.


\begin{chapthebibliography}{1}

\bibitem[Akiyama et al.(2002)]{Aki02} Akiyama, S., Wheeler, J. C.,  
   Meier, D. L. \& Lichtenstadt, I. 2002, ApJ, in press (astro-ph/0208128)
\bibitem[Aloy et al.(1999)]{Aloy99} Aloy, M.~A., Ib{\' a}{\~n}ez, J.~M.,
   Mart{\' i}, J.~M., G{\' o}mez, J.-L., \& M{\" u}ller, E.\ 1999, \apjl, 
   523, L125
\bibitem[Balbus \& Hawley(1991)]{BH91} Balbus, S. A. \& Hawley, J. F.
   1991, \apj, 376, 214
\bibitem[Balbus \& Hawley(1998)]{BH98} Balbus, S. A. \& Hawley, J. F.
   1998, Review of Modern Physics, 70, 1
\bibitem[Bisnovatyi-Kogan(1971)]{bis71} Bisnovatyi-Kogan, G. S. 1971,
   Soviet Astronomy AJ, 14, 652
\bibitem[Bisnovatyi-Kogan \& Ruzmaikin(1976)]{bis76} Bisnovatyi-Kogan, G.
   S., \& Ruzmaikin, A. A. 1976, \apss, 42, 401
\bibitem[Blandford \& Payne(1982)]{blan82} Blandford, R. D., \& Payne, D. 
   G. 1982, \mnras, 199, 883
\bibitem[Dubner et al.(2002)]{dubner02} Dubner, G., Giacani, E., Gaensler,
   B. M., Goss, W. M. \& Green. 2002, in Neutron Stars in Supernova
   Remnants, eds. P. O. Slane \& B. M. Gaensler (Astron. Soc. Pac.:
   Provo), in press (astro-ph/0112155)
\bibitem[Fesen \& Gunderson(1996)]{fesen96} Fesen, R. A., \& Gunderson, K.
   S. 1996, \apj, 470, 967
\bibitem[Fesen(2001)]{fesen01} Fesen, R. A. 2001 \apjs, 133, 161
\bibitem[Fryer \& Heger(2000)]{FH00} Fryer, C. L. \& Heger, A. 2000, \apj,
   541, 1033
\bibitem[Goussard et al. 1998]{gou98} Goussard, J-O., Haensel, P., \&
   Zdunik, J. L. 1998, \aap, 330, 1005
\bibitem[Hawley et al.(1996)]{HGB96} Hawley, J. F., Gammie, C. F., \& 
   Balbus, S. A. 1996, \apj, 464, 690
\bibitem[Hawley \& Balbus(2002)]{HB02} Hawley, J. F. \& Balbus, S. A. 
   2002, \apj, in press (astro-ph/0203309)
\bibitem[Heger et al.(2000)]{HLW00} Heger, A., Langer, N., \&
   Woosley, S. E. 2000, \apj, 528, 368
\bibitem[Heger \& Woosley(2002)]{HW02} Heger, A. \& Woosley, S. E. 2002,  
   astro-ph/0206005
\bibitem[Helfand, Gotthelf, \& Halpern(2001)]{helfand01} Helfand, D. J.,   
   Gotthelf, E. V., \& Halpern, J. P. 2001, \apj, 556, 380
\bibitem[H\"{o}flich, Khokhlov, \& Wang(2001)]{hoflich01} H\"{o}flich, P.,
   Khokhlov, A., \& Wang, L. 2001, in Proc. of the 20th Texas
   Symposium on Relativistic Astrophysics, eds. J. C. Wheeler \& H.
   Martel (New York: AIP), 459
\bibitem[Hughes et al.(2000)]{hughes00} Hughes, J. P., Rakowski, C. E.,
   Burrows, D. N., \& Slane, P. O. 2000, \apjl, 528, L109
\bibitem[Hwang et al.(2000)]{hwang00} Hwang, U., Holt, S. S., \& Petre, R.
   2000, \apjl, 537, L119
    \& Meyer-Hofmeister, E., \& Thomas, H. C. 1970, \aap, 5, 155
\bibitem[Khokhlov et al.(1999)]{khokhlov99} Khokhlov, A. M., H\"{o}flich,
   P., Oran, E. S., Wheeler, J. C., Wang, L., \& Chtchelkanova, A. Yu.
   1999, \apjl, 524, L107
\bibitem[Khokhlov \& H\"{o}flich(2001)]{khokhlov01} Khokhlov, A., \&  
   H\"{o}flich, P. 2001, in AIP Conf. Proc. No. 556, Explosive Phenomena 
   in Astrophysical Compact Objects, eds. H.-Y, Chang, C.-H., Lee, \& M. 
   Rho (New York: AIP), 301
\bibitem[Koide et al.(2000)]{koide00} Koide, S., Meier, D. L., Shibata,
   L., \& Kudoh, T. 2000, \apj, 536, 668
\bibitem[Kundt(1976)]{kundt76} Kundt, W. 1976, \nat, 261, 673
\bibitem[Laming(1999)]{laming99} Laming, J. M. 1999, New Astronomy, 4, 389
\bibitem[Leonard et al.(2000)]{leonard00} Leonard, D. C., Filippenko, A.
   V., Barth, A. J., \& Matheson, T. 2000, \apj, 536, 239
\bibitem[Leonard et al.(2001)]{leonard01} Leonard, D. C., Filippenko, A.
   V., Ardila, D. R., \& Brotherton, M. S. 2001, \apj, 553, 861
\bibitem[Li(2002)]{li02} Li, L.\ 2002, \apj, 564, 108
\bibitem[MacFayden \& Woosley(1999)]{MacF99} MacFadyen, A. \& Woosley, S. 
   E. 1999, ApJ,  524, 262
\bibitem[MacFayden, Woosley \& Heger (2001)]{MacFWH01} MacFadyen, A.,   
   Woosley, S. E., \& Heger, A. 2001, ApJ, 550, 410
\bibitem[Matzner(2002)]{Matz02} Matzner, C. D. 2002, \mnras, in press
   (astro-ph/0203085)
\bibitem[Medvedev \& Loeb]{Med99} Medvedev, M. V. \& Loeb, A. 1999, ApJ,
   526, 697
\bibitem[Meier(1999)]{meier99} Meier, D. L. 1999, \apj, 522, 753
\bibitem[Meier et al.(2001)]{meier01} Meier, D. L., Koide, S., \&, Uchida,
   Y. 2001, Science, 291, 84
\bibitem[M\'esz\'aros \& Rees(2001)]{MezRees01} M\'esz\'aros, P. \& Rees, 
   M. J. 2001, ApJ, 556, L37 
\bibitem[M\"{o}nchmeyer \& M\"{u}ller(1989)]{MM89} M\"{o}nchmeyer, R. \&
   M\"{u}ller, E. 1989, in Timing Neutron Stars, ed. H. \"{O}gelman \&  
   E. P. J. van den Heuvel (NATO ASI Ser. C, 262; New Tork: ASI), 549
\bibitem[Myra et al.(1987)]{itamar87} Myra, E. S., Bludman, S. A.,
   Hoffman, Y., Lichtenstadt, I., Sack, N., \& van Riper, K. A. 1987,
   \apj, 318, 744
\bibitem[Ogilvie(2001)]{Ogil01} Ogilvie, G.~I.\ 2001, \mnras, 325, 231
\bibitem[Ostriker \& Gunn(1971)]{ostriker71} Ostriker, J. P., \& Gunn, J.
   E. 1971, \apjl, 164, L95
\bibitem[Pun et al.(2001)]{pun01} Pun, C. S. J., \& Supernova INtensive
   Studies (SINS) Collaboration. 2001, AAS, 199, 9402
\bibitem[Ramirez Ruiz et al.(2002)]{RamR02} Ramirez Ruiz, E., Celotti, A.
   \& Rees, M. J. 2002, \mnras, in press (astro-ph/0205108)
\bibitem[Ruderman et al.(2000)]{rud00} Ruderman, M. A., Tao, L., \&
   Klu\'{z}niak, W. 2000, \apj, 542, 243
\bibitem[Spruit(2002)]{spruit02} Spruit, H. C. 2002, \aap, 381, 923
\bibitem[Spruit \& Phinney(1998)]{spruit98} Spruit, H.C., \& Phinney, E.
   S. 1998, \nat, 393, 139  
\bibitem[Spruit et al.(2001)]{spruit01} Spruit, H. C., Daigne, F., \&
   Drenkhahn, G. 2001, \aap, 369, 694
\bibitem[Tan et al.(2001)]{Tan01} Tan, J.~C., Matzner, C.~D., \& McKee,   
   C.~F.\ 2001, ApJ, 551, 946
\bibitem[Thompson \& Duncan 1996]{TD96} Thompson, C.~\&
   Duncan, R.~C.\ 1996, \apj, 473, 322
\bibitem[Thompson \& Murray(2002)]{thompson02} Thompson, C., \& Murray, N.
   2002, preprint (astro-ph/0105425)
\bibitem[Vishniac \& Cho(2001)]{VC01} Vishniac, E.~T.~\&
   Cho, J.\ 2001, \apj, 550, 752
\bibitem[Wang et al.(1996)]{wang96} Wang, L., Wheeler, J. C., Li, Z. W., 
   \& Clocchiatti, A. 1996, \apj, 467, 435
\bibitem[Wang et al.(2001)]{wang01} Wang, L., Howell, D. A.,  
   H\"{o}flich, P., \& Wheeler, J. C. 2001, \apj, 550, 1030
\bibitem[Wang et al.(2002)]{wang02} Wang, L., Weeler, J.C., H"{o}flich, 
   P., Khokhlov, A., Baade, D., Branch, D., Challis, P., Filippenko, A.V., 
   Fransson, C., Garnavich, P., Kirshner, R.P., Lundqvist, P., McCray, R., 
   Panaja, N., Pun, C.S.J., Phillips, M.M., Sonneborn, G., Suntzeff, N.B. 
   2002, \apj, in press (astro-ph/0205337) 
\bibitem[Weisskopf et al.(2000)]{weisskopf00} Weisskopf, M. C., Hester, J.
   J., Tennant, A. F., Elsner, R. F., Schulz, N. S., Marshall, H. L.,
   Karovska, M., Nicholas, J. S., Swartz, D. A., Kolodziejczak, J. J., \&
   O'Dell, S. L. 2000, \apjl, 536, L81
\bibitem[Wheeler et al.(2000)]{wheel00} Wheeler, J. C., Yi, I.,
   H\"{o}flich, P., Wang, L. 2000, \apj, 537, 810
\bibitem[Wheeler, Meier, \& Wilson(2002)]{wheel02} Wheeler, J. C., Meier,
   D. L., \& Wilson, J. R. 2002, \apj, 568, 807
\bibitem[Williams(2002)]{williams02} Williams, P. T., IAOC Workshop
   "Galactic Star Formation Across the Stellar Mass Spectrum," ASP 
   Conference Series, ed. J. M. De Buizer, in press (astro-ph/0206230)
\bibitem[Woosley \& Heger(2001)]{woo01} Woosley, S. E., \& Heger, A.
   2001, AAS, 198, 3801
\bibitem[Willingale et al.(2002)]{willingale02} Willingale, R., Bleeker,
   J. A. M., van der Heyden, K. J., Kaastra, J. S. \& Vink, J. 2002, \aap,
   381, 1039
\bibitem[Yamada \& Sato(1994)]{yamada94} Yamada, S., \& Sato, K. 1994,
   \apj, 434, 268
\bibitem[Zhang, Woosley \& MacFadyen(2002)]{zhang02} Zhang, W.,
Woosley, S. E. \& MacFadyen, A. I. 2002, ApJ, in press (astro-ph/0207436)

\end{chapthebibliography}
\end{document}